\documentclass[aps,preprint]{revtex4}%
\usepackage{amsfonts}
\usepackage{amsmath}
\usepackage{amssymb}
\usepackage{subfigure}
\usepackage{graphicx}%
\begin{document}
\preprint{CTP-SCU/2019009}
\title{Minimal Length Effect on Thermodynamics and Weak Cosmic Censorship Conjecture
in anti-de Sitter Black Holes via Charged Particle Absorption}
\author{Benrong Mu$^{a,b}$}
\email{benrongmu@cdutcm.edu.cn}
\author{Jun Tao$^{b}$}
\email{taojun@scu.edu.cn}
\author{Peng Wang$^{b}$}
\email{pengw@scu.edu.cn}
\affiliation{$^{a}$Physics Teaching and Research section, College of Medical Technology,
Chengdu University of Traditional Chinese Medicine ,Chengdu, 611137, PR China}
\affiliation{$^{b}$Center for Theoretical Physics, College of Physical Science and
Technology, Sichuan University, Chengdu, 610064, PR China}

\begin{abstract}
In this paper, we investigate minimal length effects on the thermodynamics and
weak cosmic censorship conjecture in a RN-AdS black bole via charged particle
absorption. We first use the generalized uncertainty principle (GUP) to
investigate the minimal length effect on the Hamilton-Jacobi equation. After
the deformed Hamilton-Jacobi equation is derived, we use it to study the
variations of the thermodynamic quantities of a RN-Ads black hole via
absorbing a charged practice. Furthermore, we check the second law of
thermodynamics and the weak cosmic censorship conjecture in two phase spaces.
In the normal phase space, the second law of thermodynamics and the weak
cosmic censorship conjecture are satisfied in the usual and GUP deformed
cases, and the minimal length effect makes the increase of entropy faster than
the usual case. After the charge particle absorption, the extremal RN-AdS
black hole becomes non-extremal. In the extended phase space, the black hole
entropy can either increase or decrease. When $T>2Pr_{+}$, the second law is
satisfied. When $T<2Pr_{+}$, the second law of thermodynamics is violated for
the extremal or near-extremal black hole. Finally, we find that the weak
cosmic censorship conjecture is legal for extremal and near-extremal RN-Ads
black holes in the GUP deformed case.

\end{abstract}
\keywords{}\maketitle
\tableofcontents



\section{Introduction}

\label{Sec:IN} The classical theory of black holes predicts that nothing,
including light, could escape from the black holes. However, Stephen Hawking
first showed that quantum effects could allow black holes to emit particles
\cite{INT-Hawking:1974sw}. Since then, people have begun to study the
thermodynamic properties of black holes as a thermodynamic system, and have
made a lot of achievements. For example, using the semiclassical method, Kraus
and Wilczek have modeled Hawking radiation as a tunneling effect
\cite{INT-Kraus:1994by,INT-Kraus:1994fj}. Hawking radiation related problems
are studied in depth by using the null geodesic method and the Hamilton-Jacobi
method
\cite{INT-Hemming:2001we,INT-Medved:2002zj,INT-Vagenas:2001rm,INT-Arzano:2005rs,INT-Wu:2006pz,INT-Li:2017pyf}%
. Analogous to the four laws of thermodynamics, Bardeen et al. proposed four
laws for black holes \cite{INT-Bardeen:1973gs}. The research on black holes
and gravitational waves also has made important progress. The first
gravitational wave signal GW150914 was directly detected on September 14,
2015. The signal confirms an important prediction of general relativity that
there are binary black hole systems in the universe, and they could combine to
form a larger black hole \cite{INT-Abbott:2016blz}.

General relativity predicts that the final product of the gravitational
collapse of a star could lead to a singularity of spacetime. To avoid
destructions caused by the singularity, Penrose first proposed the weak cosmic
censorship conjecture (WCCC), which states that naked singularities cannot be
formed in a real physical process from regular initial conditions
\cite{INT-WCCC-Penrose:1969pc}. In other words, the singularity is always
hidden behind the horizon, and an observer at the infinite distance can never
observe the existence of the singularity. To test the validity of the weak
cosmic censorship conjecture, Wald first tried to overcharge/overspin an
extremal Kerr-Newman black hole by throwing a test particle with
charge/angular momentum into the event horizon \cite{INT-WCCC-Wald}. It was
found that near-extremal charged/rotating black holes could be
overcharged/overspun by absorbing the test particle with charge/angular
momentum
\cite{INT-WCCC-Hubeny:1998ga,INT-WCCC-Jacobson:2009kt,INT-WCCC-Saa:2011wq,INT-WCCC-Isoyama:2011ea}%
. However, considering the back reaction and self-force effects, the study
suggests that the weak cosmic censorship conjecture may be still satisfied
\cite{INT-WCCC-Hod:2008zza,INT-WCCC-Barausse:2010ka,INT-WCCC-Barausse:2011vx,INT-WCCC-Zimmerman:2012zu,INT-WCCC-Colleoni:2015afa,INT-WCCC-Colleoni:2015ena}%
. Since there is a lack of universal evidence for the weak cosmic censorship
conjecture, its validity has been tested in various black holes
\cite{INT-WCCC-Toth:2011ab,INT-WCCC-Yu:2018eqq,INT-WCCC-Liang:2018wzd,INT-WCCC-Gwak:2018akg,INT-WCCC-Chen:2018yah,INT-WCCC-Chen:2019nsr,INT-WCCC-Han:2019kjr,INT-WCCC-Zeng:2019aao}%
. Recently, the validity of the weak cosmic censorship conjecture through
absorption of charged particles has been tested in the extended phase space,
where the cosmological constant is treated as a thermodynamic variable. The
results showed that the first law of thermodynamics and the weak cosmic
censorship conjecture are satisfied, while the second law of thermodynamics is
violated for the extremal and near-extremal black holes
\cite{INT-WCCC-Gwak:2017kkt,INT-WCCC-Chen:2019pdj,INT-WCCC-Wang:2019dzl,INT-WCCC-Hong:2019yiz,INT-WCCC-Zeng:2019hux,INT-WCCC-Zeng:2019jta,INT-WCCC-Gwak:2019rcz,INT-WCCC-Gwak:2019asi,INT-WCCC-He:2019fti,INT-WCCC-Zeng:2019huf}%
.

Since the singularity is a point where general relativity fails, we need a
broader theory to describe the gravity and quantum behavior of black holes,
especially the singularity of spacetime. On the other hand, various theories
of quantum gravity, such as loop quantum gravity, string theory, quantum
geometry and Doubly Special Relativity, imply the existence of a minimal
observable length
\cite{INT-GUP-Townsend:1977xw,INT-GUP-Freese:2002yp,INT-GUP-Das:2009qb,INT-GUP-Konishi:1989wk,INT-GUP-Padmanabhan:2015vma}%
. The generalized uncertainty principle (GUP) \cite{INT-GUP-Kempf:1994su} is
one of simple models to realize this minimal observable length. The GUP can be
derived from the deformed fundamental commutation relation
\cite{INT-GUP-Maggiore:1993kv}:
\begin{equation}
\lbrack X,P]=i\hbar\left(  1+\beta P^{2}\right)  ,
\label{deformed fundamental commutation relation}%
\end{equation}
where $\beta=\beta_{0}/m_{p}^{2}$ is the deformation parameter, $\beta_{0}$ is
a dimensionless number, and $m_{p}$ is the planck mass. The minimal observable
length is $\Delta_{\min}=\hbar\sqrt{\beta}$. The value range of $\beta_{0}$ is
constrainted as $1\lesssim\beta_{0}<10^{36}$
\cite{INT-GUP-Das:2008kaa,INT-GUP-Mu:2009rx}. For a review of GUP, see
\cite{INT-GUP-Hossenfelder:2012jw}. GUP is one of the simplest models of
effective quantum gravity, and many interesting results in the study of black
hole physics have been produced
\cite{INT-GUP2-Scardigli:1999jh,INT-GUP2-Custodio:2003jp,INT-GUP2-AmelinoCamelia:2005ik,INT-GUP2-Kim:2007nh,INT-GUP2-Ma:2013msa,INT-GUP2-Mu:2015qta,INT-GUP2-Benrong:2014woa,INT-GUP2-Chen:2016ftz,INT-GUP2-Haldar:2017viz,INT-GUP2-Ong:2018syk,INT-GUP2-Ong:2018xna,INT-GUP2-Hassanabadi:2019eol,INT-GUP-Faizal:2018fmz,INT-GUP-Faizal:2017dlb,INT-GUP-Faizal:2014mba,INT-GUP-Nowakowski:2009ha,INT-GUP-Khodadi:2018wed,INT-GUP-Khodadi:2017eim}%
. Specifically, the authors of \cite{INT-RB-Gim:2018axz} discussed the effect
of quantum gravity on the weak cosmic censorship conjecture and showed that
the second law of thermodynamics and the cosmic censorship conjecture are
violated owing to the rainbow effect.

In this paper, we will discuss the effects of quantum gravity on black hole
thermodynamics and the weak cosmic censorship conjecture in the framework of
GUP. The rest of this paper is organized as follows. In section \ref{Sec:DHJ},
we derive the GUP deformed Hamilton-Jacobi equation for a particle in the
RN-Ads spacetime and discuss its motion around the black hole horizon. In
section \ref{Sec:QET}, the minimal length effect on the thermodynamics of the
black hole is discussed in the extended phase space. In section \ref{Sec:WCCC}%
, we investigate the minimal length effect on the valitity of the weak cosmic
censorship conjecture. We summarize our results in section \ref{Sec:Con}. For
simplicity, we set $G=\hbar=c=k_{B}=1$ in this paper.

\section{Deformed Hamilton-Jacobi Equation in a RN-AdS Black Hole}

\label{Sec:DHJ}

In this section, we first review the thermodynamic properties of RN-AdS black
holes. Then, the GUP deformed Hamilton-Jacobi equation is derived, and the
motion of a charge particle near the horizon of the black hole is discussed.

The metric of a Reissner-Nordstr\"{o}m anti-de Sitter (RN-AdS) black hole in
$\left(  3+1\right)  $ curved spacetime is given by
\begin{equation}
ds^{2}=-h\left(  r\right)  dt^{2}+\frac{1}{h\left(  r\right)  }dr^{2}%
+r^{2}\left(  d\theta^{2}+\sin^{2}\theta d\varphi^{2}\right)  ,
\label{ADS-RN-metric}%
\end{equation}
with the electromagnetic potential
\begin{equation}
A_{\mu}=\left(  -\frac{Q}{r},0,0,0\right)  , \label{electromagnetic potential}%
\end{equation}
where%
\begin{equation}
h\left(  r\right)  =1-\frac{2M}{r}+\frac{Q^{2}}{r^{2}}+\frac{r^{2}}{l^{2}%
}\text{,}%
\end{equation}
\ $l$ is the AdS radius, and $M$ and $Q$ are the ADM mass and charge of the
black hole, respectively. The AdS radius $l$ is related to the cosmological
constant as $\Lambda=-3/l^{2}$. When the black hole is non-extremal, the
equation $h\left(  r\right)  =0$ has two positive real roots $r_{+}$ and
$r_{-}$, where the maximum root $r_{+}$ represents the radius of the event
horizon. When the black hole is extremal, $h\left(  r\right)  =0$ only
possesses a single root $r_{+}$. The mass of the RN-AdS black hole can be
expressed in terms of $r_{+}$
\begin{equation}
M=\frac{1}{2}\left[  r_{+}+\frac{r_{+}^{3}}{l^{2}}+\frac{Q^{2}}{r_{+}}\right]
. \label{MR}%
\end{equation}
The Hawking temperature of the AdS-RN black hole is given by
\begin{equation}
T=\frac{h^{\prime}\left(  r_{+}\right)  }{4\pi}=\frac{1}{4\pi r_{+}}\left(
1+\frac{3r_{+}^{2}}{l^{2}}-\frac{Q^{2}}{r_{+}^{2}}\right)  .
\label{Hawking temperature}%
\end{equation}
Moreover, the Bekenstein--Hawking entropy and the electric potential are
\begin{equation}
S=\frac{A}{4}=\pi r_{+}^{2}\text{ and }\Phi=\frac{Q}{r_{+}}, \label{Phi}%
\end{equation}
respectively, where $A=4\pi r_{+}^{2}$ is the horizon area.

As a stable thermodynamic system, black holes can be discussed in two phase
spaces. In the normal phase where the cosmological constant\ is a constant,
the state parameters satisfy the first law of thermodynamics
\begin{equation}
dM=TdS+\Phi dQ. \label{the first law1}%
\end{equation}
Hovever, in contrast to the usual first law of thermodynamics, the $VdP$ term
is missing in the eqn. $\left(  \ref{the first law1}\right)  $. Inspired by
this, the cosmological constant can been taken as the pressure of the black
hole \cite{DHE-1Dolan:2011xt,DHE-1Gunasekaran:2012dq}. The expression between
the cosmological constant and the pressure is given as follows
\begin{equation}
P\equiv-\frac{\Lambda}{8\pi}=\frac{3}{8\pi l^{2}}. \label{P}%
\end{equation}
The first law of thermodynamics in the extended phase space is
\begin{equation}
dM=TdS+\Phi dQ+VdP, \label{the first law}%
\end{equation}
where the valume is given by
\begin{equation}
V=\frac{4}{3}\pi r_{+}^{3}\text{.}%
\end{equation}
The mass of the black hole $M$ is defined as its enthalpy
\cite{DHE-1Cvetic:2010jb,DHE-1Kastor:2009wy}
\begin{equation}
M=U+PV. \label{enthalpy}%
\end{equation}

In \cite{INT-GUP2-Benrong:2014woa}, we have already derived the
Hamilton-Jacobi equations for a scalar particle and a fermion in a curved
spacetime background under an electric potential $A_{%
\mu
}$ and showed that these Hamilton-Jacobi equations have the same form:
\begin{equation}
\left(  \partial^{\mu}I-qA^{\mu}\right)  \left(  \partial_{\nu}I-qA_{\nu
}\right)  +m^{2}=0, \label{HJ-EQ}%
\end{equation}
where $I$ is the action, $A_{\mu}$ is the electromagnetic potential, $m$ and
$q$ are the mass and the charge of particle, respectively. In the RN-AdS
metric $\left(  \ref{ADS-RN-metric}\right)  $, the Hamilton-Jacobi equation
$\left(  \ref{HJ-EQ}\right)  $ reduces to
\begin{equation}
-\frac{1}{h(r)}\left(  \frac{\partial I}{\partial t}-qA_{t}\right)
^{2}+h(r)\left(  \partial_{r}I\right)  ^{2}+\frac{\left(  \partial_{\theta
}I\right)  ^{2}}{r^{2}}+\frac{\left(  \partial_{\phi}I\right)  ^{2}}{r^{2}%
\sin^{2}\theta}+m^{2}=0. \label{HJ-EQ2}%
\end{equation}
Taking into account the symmetries of the spacetime, we can employ the
following ansatz
\begin{equation}
I=-Et+W(r,\theta)+P_{\phi}\phi, \label{action}%
\end{equation}
where $E$ and $P_{\phi}$ are the energy and the $z$-component of angular
momentum of emitted particles, respectively. The magnitude of the angular
momentum of the particle $L$ can be expressed in terms of
\begin{equation}
L^{2}=P_{\theta}^{2}+\frac{P_{\phi}^{2}}{\sin^{2}\theta},
\label{angular momentum}%
\end{equation}
where $P_{\theta}=\partial_{\theta}W$. Plugging eqns. $\left(  \ref{action}%
\right)  $ and $\left(  \ref{angular momentum}\right)  $ into eqn. $\left(
\ref{HJ-EQ2}\right)  $, we get
\begin{equation}
E=q\Phi+\sqrt{\left[  P^{r}\left(  r_{+}\right)  \right]  ^{2}+\left(
m^{2}+\frac{L^{2}}{r^{2}}\right)  h(r)}, \label{E1}%
\end{equation}
where $P^{r}(r)=h(r)P_{r}(r)=h(r)\partial_{r}W$ is the radial momentum of the
particle. Since the energy of the particle is required to be a positive value
\cite{DHE-2Christodoulou:1970wf,DHE-2Christodoulou:1972kt}, we choose the
positive sign in front of the square root. $P^{r}$ is finite and nonzero at
event horizon $r=r_{+}$, which accounts for the Hawing radiation modeled as a
tunneling process \cite{DHE-2Pellicer:1969cf}. At the horizon $r=r_{+}$, the
eqn. $\left(  \ref{E1}\right)  $ reduces to
\begin{equation}
E=q\Phi+\left\vert P^{r}\left(  r_{+}\right)  \right\vert , \label{E2}%
\end{equation}
\ \ \ \ \ which relates the energy of the particle to its momentum and
potential energy near event horizon $r=r_{+}$.

To implement deformed fundamental commutation relation $\left(
\ref{deformed fundamental commutation relation}\right)  $, one defines%
\begin{align}
X_{i}  &  =x_{i},\nonumber\\
P_{i}  &  =p_{i}\left(  1+\beta p^{2}\right)  =p_{i}f\left(  \beta
p^{2}\right)  , \label{X-P}%
\end{align}
where $p^{2}=\sum_{i}p_{i}p_{i}$, and $f(x)=1+x$ for the Brau reduction
\cite{DHJ-1-Brau:1999uv,DHJ-1-Guo:2015ldd}. The operators $x_{i}$ and $p_{i}$
are the conventional momentum and position operators satisfying%
\begin{align}
\left[  x_{i},p_{j}\right]   &  =i\hbar\delta_{ij},\left[  x_{i},x_{j}\right]
=\left[  p_{i},p_{j}\right]  =0,\nonumber\\
x_{i}  &  =x_{i}\text{ and }p_{i}=\frac{\hbar}{i}\frac{\partial}{\partial
x_{i}}. \label{X-P2}%
\end{align}
Using the WKB method, the deformed Hamilton-Jacobi equation in the RN-AdS
metric has been obtained
\cite{DHJ-1-Guo:2015ldd,DHJ-1-Kempf:1996nk,DHJ-1-Berger:2010pj}
\begin{equation}
\frac{1}{h(r)}\left(  \frac{\partial I}{\partial t}-qA_{t}\right)  ^{2}-\chi
f^{2}(\beta\chi)=m^{2}, \label{deformed Hamilton-Jacobi equation}%
\end{equation}
where%
\begin{equation}
\chi=h(r)\left(  \partial_{r}I\right)  ^{2}+\frac{\left(  \partial_{\theta
}I\right)  ^{2}}{r^{2}}+\frac{\left(  \partial_{\phi}I\right)  ^{2}}{r^{2}%
\sin^{2}\theta}\text{.} \label{X}%
\end{equation}
To solve the deformed Hamilton-Jacobi equation $\left(
\ref{deformed Hamilton-Jacobi equation}\right)  $, we plug the ansatz eqn.
$\left(  \ref{action}\right)  $ into eqn. $\left(
\ref{deformed Hamilton-Jacobi equation}\right)  $ and find that%

\begin{equation}
\left(  E+qA_{t}\right)  ^{2}=h(r)\chi f^{2}(\beta\chi)+m^{2}h(r),
\label{deformed Hamilton-Jacobi equation 2}%
\end{equation}
where \
\begin{equation}
\chi=\frac{\left[  P^{r}\left(  r\right)  \right]  ^{2}}{h(r)}+\frac{L^{2}%
}{r^{2}}. \label{X2}%
\end{equation}

To relate $E$ to $P^{r}\left(  r\right)  $, we might want to evaluate eqn.
$\left(  \ref{deformed Hamilton-Jacobi equation 2}\right)  $ at the horiozn
$r=r_{+}$. However, GUP is an effective model, which is untrustworthy around
the Planck scale. The degrees of freedom within a few Planck lengths away from
the horizon is usually transplanckian. Moreover, the effective number of these
degrees of freedom is very small. Therefore, we evaluate eqn. $\left(
\ref{deformed Hamilton-Jacobi equation 2}\right)  $ at the stretched horizon
located at $r=r_{+}+m_{p}$ instead of the horizon. We then find that
\begin{equation}
E=P^{r}\left(  r_{+}\right)  f(\frac{\beta_{0}}{4\pi T}\frac{\left[
P^{r}\left(  r_{+}\right)  \right]  ^{2}}{m_{p}})+q\Phi. \label{E11}%
\end{equation}
where we use eqn. $\left(  \ref{Hawking temperature}\right)  $ to express
$h^{\prime}\left(  r_{+}\right)  $ in terms of $T$. When $E<q\Phi,$ the total
energy of the RN-AdS black hole flows out the event horizon, and it means the
superaddition has happened. When $E\geq q\Phi,$ the total energy of the RN-AdS
black hole flows in the event horizon. In this paper, we will study the
minimal length effect on the thermodynamics of RN-AdS black holes in the
$E\geq q\Phi$ case. When $\beta\rightarrow0,$ $f(\frac{\beta_{0}}{4\pi T}%
\frac{\left[  P^{r}\left(  r_{+}\right)  \right]  ^{2}}{m_{p}})\rightarrow1$
and eqn. $\left(  \ref{E11}\right)  $ reduces to eqn. $\left(  \ref{E2}%
\right)  $.

\section{Minimal Length Effect on Thermodynamics of a RN-AdS Black Hole via
Charged Particle Absorption}

\label{Sec:QET}

In this section, we use eqn. $\left(  \ref{E11}\right)  \ $to investigate
minimal length effects on thermodynamics of a RN-AdS black hole in the normal
and extended phase spaces via charged particle absorption. For the convenience
of the later discussion, we first give the following formulas:%

\begin{align}
\frac{\partial h}{\partial r}|_{r=r_{+}}  &  =4\pi T,\nonumber\\
\frac{\partial h}{\partial M}|_{r=r_{+}}  &  =-\frac{2}{r_{+}},\nonumber\\
\frac{\partial h}{\partial Q}|_{r=r_{+}}  &  =\frac{2Q}{r_{+}^{2}}=\frac
{2\Phi}{r_{+}},\label{partial derivative 1}\\
\frac{\partial h}{\partial P}|_{r=r_{+}}  &  =\frac{8\pi r_{+}^{2}}{3}%
=\frac{2V}{r_{+}}.\nonumber
\end{align}

\subsection{ Normal Phase Space}

When the black hole absorbs a charged particle with the mass $m$, the energy
$E$ and the charge $q$, the mass $M$ and charge $Q$ of the black hole are
varied due to the conservation law. Other thermodynamic variables of the
RN-AdS black bole would change accordingly. To check whether the changes of
the RN-AdS black hole thermodynamic variables obey the second law of
thermodynamics in the normal phase space. The initial and final state of the
RN-AdS black hole are represented by $(M,Q,r_{+})$ and $(M+dM,Q+dQ,r_{+}%
+dr_{+})$, respectively, where $dM,$ $dQ\ $and $dr_{+}$ denote the increases
of the mass, charge and radius of the RN-AdS black hole. The functions
$h(M+dM,Q+dQ,r_{+}+dr_{+})$ and $h(M,Q,r_{+})$ satisfy%
\begin{equation}
h(M+dM,Q+dQ,r_{+}+dr_{+})=h(M,Q,r_{+})+\frac{\partial h}{\partial M}%
|_{r=r_{+}}dM+\frac{\partial h}{\partial Q}|_{r=r_{+}}dQ+\frac{\partial
h}{\partial r}|_{r=r_{+}}dr_{+}, \label{hplus 1}%
\end{equation}
In the normal phase space, the black hole mass can be regarded as the internal
energy. After the black hole absorbs the particle at the event horizon, the
change of the internal energy and charge of the black hole satisfies the
following relation:
\begin{equation}
dM=E\text{ and }dQ=q. \label{change 1}%
\end{equation}
For a test particle, it's assumed that its energy $E$ and charge $q$ are small
compared to the corresponding physical quantity of the black hole,
\begin{equation}
q=dQ\ll Q\text{ and }E=dM\ll U. \label{test 1}%
\end{equation}
When the black hole absorbs a charged particle, the mass $M\ $and charge $Q$
are varied. The initial outer horizon radius $r_{+}$ moves to the final outer
horizon radius $r_{+}+dr_{+}$, which leads to
\begin{equation}
h(M,Q,r_{+})=h(M+dM,Q+dQ,r_{+}+dr_{+})=0 \label{initial and final 1}%
\end{equation}
From eqns. $\left(  \ref{partial derivative 1}\right)  $, $\left(
\ref{hplus 1}\right)  $ and $\left(  \ref{initial and final 1}\right)  ,\,$the
infinitesimal changes in $r_{+},$ $M\ $and $Q$ are related by
\begin{equation}
4\pi Tdr_{+}-\frac{2}{r_{+}}E+\frac{2\Phi}{r_{+}}q=0.
\label{infinitesimal changes 1}%
\end{equation}
Combining eqn. $\left(  \ref{infinitesimal changes 1}\right)  $ with eqns.
$\left(  \ref{Phi}\right)  $ and $\left(  \ref{E11}\right)  $, we get
\begin{equation}
dS=\frac{P^{r}\left(  r_{+}\right)  f(\frac{\beta_{0}}{4\pi T}\frac{\left[
P^{r}\left(  r_{+}\right)  \right]  ^{2}}{m_{p}})}{T}>0. \label{dSL1}%
\end{equation}
Since $f(\frac{\beta_{0}}{4\pi T}\frac{\left[  P^{r}\left(  r_{+}\right)
\right]  ^{2}}{m_{p}})>0,$ the minimal length effect does not the sign of
$dS$, which shows that the second law of thermodynamics holds in the normal
phase space. For the Brau reduction with $f\left(  x\right)  =1+x$, the GUP
effect makes the increase of the entropy faster than the usual case.

\subsection{Extended Phase Space}

In the extended phase space, the cosmological constant can be treated as the
pressure of the black hole. So the initial and final state of the RN-AdS black
hole are represented by $(M,Q,P,r_{+})$ and $(M+dM,Q+dQ,P+dP,r_{+}+dr_{+})$,
respectively. The functions $h(M+dM,Q+dQ,P+dP,r_{+}+dr_{+})$ and
$h(M,Q,P,r_{+})$ satisfy%
\begin{align}
h(M+dM,Q+dQ,r_{+}+dr_{+})  &  =h(M,Q,P,r_{+})+\frac{\partial h}{\partial
M}|_{r=r_{+}}dM+\frac{\partial h}{\partial Q}|_{r=r_{+}}dQ\nonumber\\
&  +\frac{\partial h}{\partial P}|_{r=r_{+}}dP+\frac{\partial h}{\partial
r}|_{r=r_{+}}dr_{+}, \label{hplus}%
\end{align}
In the extended phase space, the black hole mass can be regarded as the
gravitational enthalpy which relates to the internal energy as eqn. $\left(
\ref{enthalpy}\right)  $. After the black hole absorbs a particle with the
energy $E$ and charge $q$ at the event horizon, the change of internal energy
and charge of the black hole satisfy
\begin{equation}
d(M-PV)=E\text{ and }dQ=q. \label{change}%
\end{equation}
The initial outer horizon radius $r_{+}$ moves to the final outer horizon
radius $r_{+}+dr_{+}$, which leads to
\begin{equation}
h(M,Q,P,r_{+})=h(M+dM,Q+dQ,P+dP,r_{+}+dr_{+})=0 \label{initial and final}%
\end{equation}
From eqns. $\left(  \ref{partial derivative 1}\right)  $, $\left(
\ref{hplus}\right)  $ and $\left(  \ref{initial and final}\right)  ,\,$the
infinitesimal changes in $r_{+},$ $M,$ $Q$ and $P$ are related by
\begin{equation}
4\pi Tdr_{+}-\frac{2}{r_{+}}E-\frac{2P}{r_{+}}dV+\frac{2\Phi}{r_{+}}q=0.
\label{infinitesimal changes}%
\end{equation}
Combining eqn. $\left(  \ref{infinitesimal changes}\right)  $ with eqns.
$\left(  \ref{Phi}\right)  $ and $\left(  \ref{E11}\right)  $, we get
\begin{equation}
dS=\frac{P^{r}\left(  r_{+}\right)  f(\frac{\beta_{0}}{4\pi T}\frac{\left[
P^{r}\left(  r_{+}\right)  \right]  ^{2}}{m_{p}})}{T-2Pr_{+}}. \label{dSL}%
\end{equation}
Since $f(\frac{\beta_{0}}{4\pi T}\frac{\left[  P^{r}\left(  r_{+}\right)
\right]  ^{2}}{m_{p}})>0,$ the minimal length effect changes the rate of $dS$
without changing the sign of $dS.$ When $T>2Pr_{+}$, the sign of the
denominator in eqn. $\left(  \ref{dSL}\right)  $ is always positive. It means
that the second law is satisfied in the extended phase space when the black
hole is far enough from extremity. When $T<2Pr_{+}$, the sign of the
denominator in eqn. $\left(  \ref{dSL}\right)  $ is always negative. It
implies that the second law of thermodynamics is not satisfied for the
extremal or near-extremal black hole.

\section{Minimal Length Effect on Weak Cosmic Censorship Conjecture in a
RN-AdS Black Hole}

\label{Sec:WCCC}

In this section, we investigate the minimal length effect on the weak cosmic
censorship conjecture by charge particle absorption in the normal and extended
phase spaces. In the test particle limit, the black hole needs to start out
close to extremal to have chance to become a naked singularity. So we assume
that the initial RN-AdS black hole is a near-extreme black hole. Between two
event horizons, there is one and only one minimum point at $r=r_{\min}$ with
$h^{\prime}(r_{\min})=0$. In the near-extremal and extremal cases, the minimum
value of $h(r)$ is not greater than zero,%
\begin{equation}
\sigma\equiv h(r_{\min})\leq0,
\end{equation}
where $\sigma\equiv0$ corresponds to the extremal case. If there is a positive
real root in the final state $h(r_{\min}+dr_{\min})=0$, the weak cosmic
censorship conjecture is valid. Otherwise, the weak cosmic censorship
conjecture is violated. Again, we first give the following partial derivative
formulas:
\begin{align}
\frac{\partial h}{\partial r}|_{r=r_{\min}}  &  =0,\nonumber\\
\frac{\partial h}{\partial M}|_{r=r_{\min}}  &  =-\frac{2}{r},\nonumber\\
\frac{\partial h}{\partial Q}|_{r=r_{\min}}  &  =\frac{2Q}{r_{+}^{2}%
},\label{infinitesimal changes 3}\\
\frac{\partial h}{\partial P}|_{r=r_{\min}}  &  =\frac{8\pi r^{2}}%
{3}.\nonumber
\end{align}

\subsection{Normal Phase Space}

After absorbing a charged particle with the mass $m$, the energy $E$ and the
charge $q$, the physical parameters of the black hole change from the initial
state $(M,Q,r_{+})$ to the final state $(M+dM,Q+dQ,r_{+}+dr_{+})$. The final
state value of $h(r)$ at $r=r_{\min}+dr_{\min}$ is given by
\begin{equation}
h(M+dM,Q+dQ,r_{\min}+dr_{\min})=\sigma+\frac{\partial h}{\partial
M}|_{r=r_{\min}}dM+\frac{\partial h}{\partial Q}|_{r=r_{\min}}dQ+\frac
{\partial h}{\partial r}|_{r=r_{\min}}dr_{\min}. \label{hplus 2}%
\end{equation}
Using eqns. $\left(  \ref{change 1}\right)  ,$ $\left(  \ref{dSL1}\right)  $
and $\left(  \ref{infinitesimal changes 3}\right)  $, eqn. $\left(
\ref{hplus 2}\right)  $ reduces to
\begin{equation}
h(M+dM,Q+dQ,r_{\min}+dr_{\min})=\sigma-\frac{P^{r}\left(  r_{+}\right)
f\left(  \frac{\beta_{0}}{4\pi T}\frac{\left[  P^{r}\left(  r_{+}\right)
\right]  ^{2}}{m_{p}}\right)  }{r_{\min}}-\frac{2Qq}{r_{\min}}\left(  \frac
{1}{r_{+}}-\frac{1}{{r_{\min}}}\right)  \label{hplus4}%
\end{equation}

When the initial black hole is extremal RN-Ads black hole, $h(r)=0$ has only
one solution in this case, $r_{+}=r_{\min}$ and $\sigma=0$. Therefore eqn.
$\left(  \ref{hplus4}\right)  $ reduces to%
\begin{equation}
h(M+dM,Q+dQ,r_{\min}+dr_{\min})=-\frac{P^{r}\left(  r_{+}\right)
f(\frac{\beta_{0}}{4\pi T}\frac{\left[  P^{r}\left(  r_{+}\right)  \right]
^{2}}{m_{p}})}{r_{\min}},
\end{equation}
which means that an extremal RN-Ads black hole becomes a non-extremal one
after the absorption of a charged particle in the normal phase space.

For a near-extremal RN-Ads black hole, by introducing infinitesimal parameters
$\varepsilon$, we can define the relationship between $r_{\min}$ and $r_{+}$
as follows:
\begin{equation}
r_{\min}=r_{+}(1-\epsilon),
\end{equation}
where $\varepsilon\ll1.$ So $\sigma$ is suppressed by $\varepsilon$ in the
near-extremal limit. Moreover, the second term of eqn. $\left(  \ref{hplus4}%
\right)  $ is only suppressed by the test particle limit, and the third term
is suppressed by both the near-extremal limit and the test particle limit, and
hence can be neglected. Therefore, eqn. $\left(  \ref{hplus4}\right)  $ leads
to
\begin{equation}
h(M+dM,Q+dQ,r_{+}+dr_{+})=\sigma-\frac{P^{r}\left(  r_{+}\right)
f(\frac{\beta_{0}}{4\pi T}\frac{\left[  P^{r}\left(  r_{+}\right)  \right]
^{2}}{m_{p}})}{r_{\min}}<0.
\end{equation}
which means that a near-extremal black hole stays near-extremal after the
absorption of a charged test particle in the normal phase space. In the normal
phase space, we find the weak cosmic censorship conjecture is legal for
extremal and near-extremal RN-Ads black holes.

\subsection{Extended Phase Space}

In this section, we investigate the minimal length effect on the weak cosmic
censorship conjecture via charged particle absorption in the extended phase
space. After absorbing a charged particle, the physical parameters of the
black hole change from the initial state $(M,Q,P,r_{+})$ to the final state
$(M+dM,Q+dQ,P+dP,r_{+}+dr_{+})$. The final state value of $h(r)$ at
$r=r_{\min}+dr_{\min}$ is given by
\begin{align}
&  h(M+dM,Q+dQ,P+dP,r_{\min}+dr_{\min})\nonumber\\
&  =\sigma+\frac{\partial h}{\partial M}|_{r=r_{\min}}dM+\frac{\partial
h}{\partial Q}|_{r=r_{\min}}dQ+\frac{\partial h}{\partial P}|_{r=r_{\min}%
}dP+\frac{\partial h}{\partial r}|_{r=r_{\min}}dr_{\min}. \label{hplus3}%
\end{align}
Using eqns. $\left(  \ref{change}\right)  ,$ $\left(  \ref{dSL}\right)  $ and
$\left(  \ref{infinitesimal changes 3}\right)  ,$ eqn. $\left(  \ref{hplus3}%
\right)  $ reduces to
\begin{align}
&  h(M+dM,Q+dQ,P+dP,r_{\min}+dr_{\min})\nonumber\\
&  =\sigma{-}\frac{2TP^{r}\left(  r_{+}\right)  f(\frac{\beta m_{p}}{4\pi
T}\left(  p^{r}\right)  ^{2})}{\left(  T-2Pr_{+}\right)  r_{\min}}-\frac
{2Qq}{r_{\min}}\left(  \frac{1}{r_{+}}-\frac{1}{{r_{\min}}}\right)
-\frac{8\pi dP}{r_{\min}}\left(  r_{+}^{3}-r_{\min}^{3}\right)  .
\label{hplus5}%
\end{align}

When the initial black hole is an extremal RN-Ads black hole, $h(r)=0$ has
only one solution. In this case, $r_{+}=r_{\min}$, $T=0$, and $\sigma=0$.
Therefore eqn. $\left(  \ref{hplus5}\right)  $ reduces to%
\begin{equation}
h(M+dM,Q+dQ,P+dP,r_{\min}+dr_{\min})=0,
\end{equation}
which means that an extremal RN-Ads black hole stays extremal state after the
absorption of a charged particle.

For a near-extremal RN-Ads black hole, the second term of eqn. $\left(
\ref{hplus5}\right)  $ is only suppressed by the test particle limit, however
the third and fourth terms of eqn. $\left(  \ref{hplus5}\right)  $ are
suppressed by both the near-extremal limit and the test particle limit, and
hence can be neglected. Therefore, eqn. $\left(  \ref{hplus5}\right)  $ leads
to
\begin{equation}
h(M+dM,Q+dQ,P+dP,r_{\min}+dr_{\min})=\sigma{-}\frac{2TP^{r}\left(
r_{+}\right)  f(\frac{\beta_{0}}{4\pi T}\frac{\left[  P^{r}\left(
r_{+}\right)  \right]  ^{2}}{m_{p}})}{\left(  T-2Pr_{+}\right)  r_{\min}}<0.
\end{equation}
which means that a near-extremal black hole stays near-extremal after the
absorption of the charged test particle. In the extended phase space, we find
the weak cosmic censorship conjecture is legal for extremal and near-extremal
RN-Ads black holes.

\section{Conclusion}

\label{Sec:Con}

In this paper, we investigated the minimal length effect on the weak cosmic
censorship conjecture in a RN-AdS black bole via the charged particle
absorption. We first introduced thermodynamics of RN-Ads black holes in the
normal and extended phase space. Then, we employed GUP to investigate the
effect of the minimal length on the Hamilton-Jacobi equation. Specifically, we
derived the deformed Hamilton-Jacobi equation which has the same form both for
scalar and fermionic particles, and used it to study the variations of the
thermodynamic quantities of a RN-Ads black hole via absorbing a charged
particle. Furthermore, we checked the second law of thermodynamics and the
weak cosmic censorship conjecture in a RN-Ads black hole. In the normal phase
space, the second law of thermodynamics is satisfied, and the GUP effect made
the increase of entropy faster than the usual case. We found that the weak
cosmic censorship conjecture is satisfied for the extremal and near-extremal
RN-AdS black holes. After the charge particle absorption, an extremal RN-AdS
black hole becomes non-extremal. In the extended phase space, where the
cosmological constant is treated as pressure, the black hole entropy can
either increase or decrease depending on the states of black hole. When
$T>2Pr_{+},$ the second law is satisfied in the extended phase space. When
$T<2Pr_{+},$ the second law of thermodynamics is not satisfied. Finally, we
found that the weak cosmic censorship conjecture is always legal for extremal
and near-extremal RN-Ads black holes.

To discuss the validity of the weak cosmic censorship, we considered the
minimum value of the metric function $h\left(  r\right)  $ of an extremal or
near-extremal black hole after a charged particle is absorbed by the black
hole. It was found that the minimum value stays negative, which means the
event horizon of the black hole always exists, and hence the weak cosmic
censorship is satisfied. However, apart from the existence of the horizon, the
analysis did not provide the information about how the position of the horizon
changes. On the other hand, the test of the second law may shed light on how
the horizon radius changes. The horizon expands when the second law is valid,
while the horizon shrinks when the second law is violated (e.g., for an
extremal or near-extremal black hole in the extended phase space). In other
words, the validities of the weak cosmic censorship and the second law
crucially depend on the existence and the position of the horizon, respectively,
after the black hole absorbs the particle.

In \cite{INT-RB-Gim:2018axz}, the second law of thermodynamics and the weak
cosmic censorship conjecture of the RN rainbow black hole have been
investigated. The authors considered only the normal phase space case.
Different from \cite{INT-RB-Gim:2018axz}, we have chosen another quantum
gravity model, namely GUP, to study the second law of thermodynamics and the
weak cosmic censorship conjecture in the normal and extended phase space. When
$\beta\rightarrow0,$ $f(\frac{\beta_{0}}{4\pi T}\frac{\left[  P^{r}\left(
r_{+}\right)  \right]  ^{2}}{m_{p}})\rightarrow1$ and the GUP deformed case
reduces to the usual case, and our result is consistent with that in
\cite{INT-WCCC-Gwak:2018akg,INT-WCCC-Chen:2019nsr,INT-WCCC-Wang:2019dzl}.

\section{\noindent\textbf{Acknowledgements }}

We are grateful to Prof. H. Yang, Prof. D. Chen and Dr. H. Wu for their useful
discussions. This work is supported in part by the key fund project for
Education Department of Sichuan (Grant no. 18ZA0173) and NSFC (Grant
No.11747171, 11875196, 11375121 and 11005016). Natural Science Foundation of
Chengdu University of TCM (Grants nos. ZRYY1921) and Discipline Talent
Promotion Program of Xinglin Scholars(Grant no. QNXZ2018050).


\begin{thebibliography}{99}                                                                                               %


\bibitem {INT-Hawking:1974sw}S.~W.~Hawking, ``Particle Creation by Black
Holes,'' Commun.\ Math.\ Phys.\ \textbf{43}, 199 (1975) Erratum:
[Commun.\ Math.\ Phys.\ \textbf{46}, 206 (1976)]. doi:10.1007/BF02345020,
10.1007/BF01608497




\bibitem {INT-Kraus:1994by}P.~Kraus and F.~Wilczek, ``Selfinteraction
correction to black hole radiance,'' Nucl.\ Phys.\ B \textbf{433}, 403 (1995)
doi:10.1016/0550-3213(94)00411-7 [gr-qc/9408003].




\bibitem {INT-Kraus:1994fj}P.~Kraus and F.~Wilczek, ``Effect of
selfinteraction on charged black hole radiance,'' Nucl.\ Phys.\ B
\textbf{437}, 231 (1995) doi:10.1016/0550-3213(94)00588-6 [hep-th/9411219].




\bibitem {INT-Hemming:2001we}S.~Hemming and E.~Keski-Vakkuri, ``The Spectrum
of strings on BTZ black holes and spectral flow in the SL(2,R) WZW model,''
Nucl.\ Phys.\ B \textbf{626}, 363 (2002) doi:10.1016/S0550-3213(02)00021-4
[hep-th/0110252].


\bibitem {INT-Medved:2002zj}A.~J.~M.~Medved, ``Radiation via tunneling from a
de Sitter cosmological horizon,'' Phys.\ Rev.\ D \textbf{66}, 124009 (2002)
doi:10.1103/PhysRevD.66.124009 [hep-th/0207247].


\bibitem {INT-Vagenas:2001rm}E.~C.~Vagenas, ``Semiclassical corrections to the
Bekenstein-Hawking entropy of the BTZ black hole via selfgravitation,''
Phys.\ Lett.\ B \textbf{533}, 302 (2002) doi:10.1016/S0370-2693(02)01695-7
[hep-th/0109108].


\bibitem {INT-Arzano:2005rs}M.~Arzano, A.~J.~M.~Medved and E.~C.~Vagenas,
``Hawking radiation as tunneling through the quantum horizon,'' JHEP
\textbf{0509}, 037 (2005) doi:10.1088/1126-6708/2005/09/037 [hep-th/0505266].


\bibitem {INT-Wu:2006pz}S.~Q.~Wu and Q.~Q.~Jiang, ``Remarks on Hawking
radiation as tunneling from the BTZ black holes,'' JHEP \textbf{0603}, 079
(2006) doi:10.1088/1126-6708/2006/03/079 [hep-th/0602033].


\bibitem {INT-Li:2017pyf}G.~Q.~Li, ``Hawking radiation and entropy of a black
hole in Lovelock-Born-Infeld gravity from the quantum tunneling approach,''
Chin.\ Phys.\ C \textbf{41}, no. 4, 045103 (2017).
doi:10.1088/1674-1137/41/4/045103




\bibitem {INT-Bardeen:1973gs}J.~M.~Bardeen, B.~Carter and S.~W.~Hawking, ``The
Four laws of black hole mechanics,'' Commun.\ Math.\ Phys.\ \textbf{31}, 161
(1973). doi:10.1007/BF01645742


\bibitem {INT-Abbott:2016blz}B.~P.~Abbott \textit{et al.} [LIGO Scientific and
Virgo Collaborations],
Phys.\ Rev.\ Lett.\ \textbf{116}, no. 6, 061102 (2016)
doi:10.1103/PhysRevLett.116.061102 [arXiv:1602.03837 [gr-qc]].


\bibitem {INT-WCCC-Penrose:1969pc}

R.~Penrose, ``Gravitational collapse: The role of general relativity,''
Riv.\ Nuovo Cim.\ \textbf{1}, 252 (1969) [Gen.\ Rel.\ Grav.\ \textbf{34}, 1141
(2002)].


\bibitem {INT-WCCC-Wald}R. Wald, \textquotedblleft Gedanken experiments to
destroy a black hole,\textquotedblright\ Ann. Phys.\textbf{ 82}, 548 (1974)
doi:10.1016/0003-4916(74)90125-0


\bibitem {INT-WCCC-Hubeny:1998ga}V.~E.~Hubeny, ``Overcharging a black hole and
cosmic censorship,'' Phys.\ Rev.\ D \textbf{59}, 064013 (1999)
doi:10.1103/PhysRevD.59.064013 [gr-qc/9808043].


\bibitem {INT-WCCC-Jacobson:2009kt}T.~Jacobson and T.~P.~Sotiriou,
``Over-spinning a black hole with a test body,''
Phys.\ Rev.\ Lett.\ \textbf{103}, 141101 (2009) Erratum:
[Phys.\ Rev.\ Lett.\ \textbf{103}, 209903 (2009)]
doi:10.1103/PhysRevLett.103.209903, 10.1103/PhysRevLett.103.141101
[arXiv:0907.4146 [gr-qc]].


\bibitem {INT-WCCC-Saa:2011wq}A.~Saa and R.~Santarelli, ``Destroying a
near-extremal Kerr-Newman black hole,'' Phys.\ Rev.\ D \textbf{84}, 027501
(2011) doi:10.1103/PhysRevD.84.027501 [arXiv:1105.3950 [gr-qc]].


\bibitem {INT-WCCC-Isoyama:2011ea}S.~Isoyama, N.~Sago and T.~Tanaka, ``Cosmic
censorship in overcharging a Reissner-Nordstr\'{o}m black hole via charged
particle absorption,'' Phys.\ Rev.\ D \textbf{84}, 124024 (2011)
doi:10.1103/PhysRevD.84.124024 [arXiv:1108.6207 [gr-qc]].




\bibitem {INT-WCCC-Hod:2008zza}S.~Hod, ``Weak Cosmic Censorship: As Strong as
Ever,'' Phys.\ Rev.\ Lett.\ \textbf{100}, 121101 (2008)
doi:10.1103/PhysRevLett.100.121101 [arXiv:0805.3873 [gr-qc]].


\bibitem {INT-WCCC-Barausse:2010ka}E.~Barausse, V.~Cardoso and G.~Khanna,
``Test bodies and naked singularities: Is the self-force the cosmic censor?,''
Phys.\ Rev.\ Lett.\ \textbf{105}, 261102 (2010)
doi:10.1103/PhysRevLett.105.261102 [arXiv:1008.5159 [gr-qc]].




\bibitem {INT-WCCC-Barausse:2011vx}E.~Barausse, V.~Cardoso and G.~Khanna,
``Testing the Cosmic Censorship Conjecture with point particles: the effect of
radiation reaction and the self-force,'' Phys.\ Rev.\ D \textbf{84}, 104006
(2011) doi:10.1103/PhysRevD.84.104006 [arXiv:1106.1692 [gr-qc]].


\bibitem {INT-WCCC-Zimmerman:2012zu}P.~Zimmerman, I.~Vega, E.~Poisson and
R.~Haas, ``Self-force as a cosmic censor,'' Phys.\ Rev.\ D \textbf{87}, no. 4,
041501 (2013) doi:10.1103/PhysRevD.87.041501 [arXiv:1211.3889 [gr-qc]].


\bibitem {INT-WCCC-Colleoni:2015afa}M.~Colleoni and L.~Barack, ``Overspinning
a Kerr black hole: the effect of self-force,'' Phys.\ Rev.\ D \textbf{91},
104024 (2015) doi:10.1103/PhysRevD.91.104024 [arXiv:1501.07330 [gr-qc]].


\bibitem {INT-WCCC-Colleoni:2015ena}M.~Colleoni, L.~Barack, A.~G.~Shah and
M.~van de Meent, ``Self-force as a cosmic censor in the Kerr overspinning
problem,'' Phys.\ Rev.\ D \textbf{92}, no. 8, 084044 (2015)
doi:10.1103/PhysRevD.92.084044 [arXiv:1508.04031 [gr-qc]].




\bibitem {INT-WCCC-Toth:2011ab}G.~Z.~Toth, ``Test of the weak cosmic
censorship conjecture with a charged scalar field and dyonic Kerr-Newman black
holes,'' Gen.\ Rel.\ Grav.\ \textbf{44}, 2019 (2012)
doi:10.1007/s10714-012-1374-z [arXiv:1112.2382 [gr-qc]].




\bibitem {INT-WCCC-Yu:2018eqq}T.~Y.~Yu and W.~Y.~Wen, ``Cosmic censorship and
Weak Gravity Conjecture in the Einstein--Maxwell-dilaton theory,''
Phys.\ Lett.\ B \textbf{781}, 713 (2018) doi:10.1016/j.physletb.2018.04.060
[arXiv:1803.07916 [gr-qc]].




\bibitem {INT-WCCC-Liang:2018wzd}B.~Liang, S.~W.~Wei and Y.~X.~Liu, ``Weak
cosmic censorship conjecture in Kerr black holes of modified gravity,''
Mod.\ Phys.\ Lett.\ A \textbf{34}, no. 05, 1950037 (2019)
doi:10.1142/S0217732319500378 [arXiv:1804.06966 [gr-qc]].




\bibitem {INT-WCCC-Gwak:2018akg}B.~Gwak, ``Weak Cosmic Censorship Conjecture
in Kerr-(Anti-)de Sitter Black Hole with Scalar Field,'' JHEP \textbf{1809},
081 (2018) doi:10.1007/JHEP09(2018)081 [arXiv:1807.10630 [gr-qc]].




\bibitem {INT-WCCC-Chen:2018yah}D.~Chen, ``Weak cosmic censorship conjecture
in BTZ black holes with scalar fields,'' Chin.\ Phys.\ C \textbf{44}, 015101
(2020) [arXiv:1812.03459 [gr-qc]].




\bibitem {INT-WCCC-Chen:2019nsr}D.~Chen, W.~Yang and X.~Zeng, ``Thermodynamics
and weak cosmic censorship conjecture in Reissner-Nordstr$\ddot{o}$m anti-de
Sitter black holes with scalar field,'' Nucl.\ Phys.\ B \textbf{946}, 114722
(2019) doi:10.1016/j.nuclphysb.2019.114722 [arXiv:1901.05140 [hep-th]].




\bibitem {INT-WCCC-Han:2019kjr}Y.~W.~Han, X.~X.~Zeng and Y.~Hong,
``Thermodynamics and weak cosmic censorship conjecture of the torus-like black
hole,'' Eur.\ Phys.\ J.\ C \textbf{79}, no. 3, 252 (2019)
doi:10.1140/epjc/s10052-019-6771-y [arXiv:1901.10660 [hep-th]].


\bibitem {INT-WCCC-Zeng:2019aao}X.~X.~Zeng and H.~Q.~Zhang, ``Thermodynamics
and weak cosmic censorship conjecture in the Kerr-AdS black hole,''
arXiv:1905.01618 [gr-qc].




\bibitem {INT-WCCC-Gwak:2017kkt}B.~Gwak, ``Thermodynamics with Pressure and
Volume under Charged Particle Absorption,'' JHEP \textbf{1711}, 129 (2017)
doi:10.1007/JHEP11(2017)129 [arXiv:1709.08665 [gr-qc]].




\bibitem {INT-WCCC-Chen:2019pdj}D.~Chen, ``Thermodynamics and weak cosmic
censorship conjecture in extended phase spaces of anti-de Sitter black holes
with particles' absorption,'' Eur.\ Phys.\ J.\ C \textbf{79}, no. 4, 353
(2019) doi:10.1140/epjc/s10052-019-6874-5 [arXiv:1902.06489 [hep-th]].




\bibitem {INT-WCCC-Wang:2019dzl}P.~Wang, H.~Wu and H.~Yang, ``Thermodynamics
of nonlinear electrodynamics black holes and the validity of weak cosmic
censorship at charged particle absorption,'' Eur.\ Phys.\ J.\ C \textbf{79},
no. 7, 572 (2019). doi:10.1140/epjc/s10052-019-7090-z




\bibitem {INT-WCCC-Hong:2019yiz}W.~Hong, B.~Mu and J.~Tao,




\bibitem {INT-WCCC-Zeng:2019hux}X.~X.~Zeng, X.~Y.~Hu and K.~J.~He, ``Weak
cosmic censorship conjecture with pressure and volume in Gauss-Bonnet AdS
black holes,'' arXiv:1905.07750 [hep-th].


\bibitem {INT-WCCC-Zeng:2019jta}X.~X.~Zeng and H.~Q.~Zhang, ``Thermodynamics
and weak cosmic censorship conjecture in Born-Infeld-anti-de Sitter black
holes,'' arXiv:1901.04247 [hep-th].




\bibitem {INT-WCCC-INT-Gwak:2019rcz}B.~Gwak, ``Weak Cosmic Censorship in
Kerr-Sen Black Hole under Charged Scalar Field,'' arXiv:1910.13329 [gr-qc].




\bibitem {INT-WCCC-Gwak:2019asi}B.~Gwak, ``Weak Cosmic Censorship with
Pressure and Volume in Charged Anti-de Sitter Black Hole under Charged Scalar
Field,'' JCAP \textbf{1908}, 016 (2019) doi:10.1088/1475-7516/2019/08/016
[arXiv:1901.05589 [gr-qc]].




\bibitem {INT-WCCC-He:2019fti}K.~J.~He, X.~Y.~Hu and X.~X.~Zeng, ``The weak
cosmic censorship conjecture and thermodynamics in the quintessence AdS black
hole under charge particle absorption,'' Chin.\ Phys.\ C \textbf{43}, no. 12,
125101 (2019) doi:10.1088/1674-1137/43/12/125101 [arXiv:1906.10531 [gr-qc]].




\bibitem {INT-WCCC-Zeng:2019huf}X.~X.~Zeng, Y.~W.~Han and D.~Y.~Chen,
``Thermodynamics and weak cosmic censorship conjecture of BTZ black holes in
extended phase space,'' Chin.\ Phys.\ C \textbf{43}, no. 10, 105104 (2019)
doi:10.1088/1674-1137/43/10/105104 [arXiv:1901.08915 [gr-qc]].




\bibitem {INT-GUP-Townsend:1977xw}P.~K.~Townsend, ``Small Scale Structure of
Space-Time as the Origin of the Gravitational Constant,'' Phys.\ Rev.\ D
\textbf{15}, 2795 (1977). doi:10.1103/PhysRevD.15.2795




\bibitem {INT-GUP-Freese:2002yp}K.~Freese, M.~Lewis and J.~P.~van der Schaar,
``Observational tests of open strings in brane world scenarios,'' JHEP
\textbf{0307}, 026 (2003) doi:10.1088/1126-6708/2003/07/026 [hep-ph/0211250].


\bibitem {INT-GUP-Das:2009qb}S.~Das, S.~Ghosh and D.~Roychowdhury,
``Relativistic Thermodynamics with an Invariant Energy Scale,'' Phys.\ Rev.\ D
\textbf{80}, 125036 (2009) doi:10.1103/PhysRevD.80.125036 [arXiv:0908.0413
[hep-th]].




\bibitem {INT-GUP-Konishi:1989wk}K.~Konishi, G.~Paffuti and P.~Provero,
``Minimum Physical Length and the Generalized Uncertainty Principle in String
Theory,'' Phys.\ Lett.\ B \textbf{234}, 276 (1990).
doi:10.1016/0370-2693(90)91927-4


\bibitem {INT-GUP-Padmanabhan:2015vma}T.~Padmanabhan, S.~Chakraborty and
D.~Kothawala,
Gen.\ Rel.\ Grav.\ \textbf{48}, no. 5, 55 (2016) doi:10.1007/s10714-016-2053-2
[arXiv:1507.05669 [gr-qc]].




\bibitem {INT-GUP-Kempf:1994su}A.~Kempf, G.~Mangano and R.~B.~Mann, ``Hilbert
space representation of the minimal length uncertainty relation,''
Phys.\ Rev.\ D \textbf{52}, 1108 (1995) doi:10.1103/PhysRevD.52.1108
[hep-th/9412167].


\bibitem {INT-GUP-Maggiore:1993kv}M.~Maggiore, ``The Algebraic structure of
the generalized uncertainty principle,'' Phys.\ Lett.\ B \textbf{319}, 83
(1993) doi:10.1016/0370-2693(93)90785-G [hep-th/9309034].




\bibitem {INT-GUP-Das:2008kaa}S.~Das and E.~C.~Vagenas, ``Universality of
Quantum Gravity Corrections,'' Phys.\ Rev.\ Lett.\ \textbf{101}, 221301 (2008)
doi:10.1103/PhysRevLett.101.221301 [arXiv:0810.5333 [hep-th]].


\bibitem {INT-GUP-Mu:2009rx}B.~Mu, H.~Wu and H.~Yang, ``The Generalized
Uncertainty Principle on the Presence of Extra Dimensions,''
Chin.\ Phys.\ Lett.\ \textbf{28}, 091101 (2011)
doi:10.1088/0256-307X/28/9/091101 [arXiv:0909.3635 [hep-th]].


\bibitem {INT-GUP-Hossenfelder:2012jw}S.~Hossenfelder, ``Minimal Length Scale
Scenarios for Quantum Gravity,'' Living Rev.\ Rel.\ \textbf{16}, 2 (2013)
doi:10.12942/lrr-2013-2 [arXiv:1203.6191 [gr-qc]].




\bibitem {INT-GUP2-Scardigli:1999jh}F.~Scardigli, ``Generalized uncertainty
principle in quantum gravity from micro - black hole Gedanken experiment,''
Phys.\ Lett.\ B \textbf{452}, 39 (1999) doi:10.1016/S0370-2693(99)00167-7
[hep-th/9904025].


\bibitem {INT-GUP2-Custodio:2003jp}P.~S.~Custodio and J.~E.~Horvath, ``The
Generalized uncertainty principle, entropy bounds and black hole
(non)evaporation in a thermal bath,'' Class.\ Quant.\ Grav.\ \textbf{20}, L197
(2003) doi:10.1088/0264-9381/20/14/103 [gr-qc/0305022].


\bibitem {INT-GUP2-AmelinoCamelia:2005ik}G.~Amelino-Camelia, M.~Arzano,
Y.~Ling and G.~Mandanici, ``Black-hole thermodynamics with modified dispersion
relations and generalized uncertainty principles,''
Class.\ Quant.\ Grav.\ \textbf{23}, 2585 (2006) doi:10.1088/0264-9381/23/7/022
[gr-qc/0506110].


\bibitem {INT-GUP2-Kim:2007nh}W.~Kim, Y.~W.~Kim and Y.~J.~Park, ``Entropy of a
charged black hole in two dimensions without cutoff,'' Phys.\ Rev.\ D
\textbf{75}, 127501 (2007) doi:10.1103/PhysRevD.75.127501 [gr-qc/0702018
[GR-QC]].




\bibitem {INT-GUP2-Ma:2013msa}M.~S.~Ma and H.~H.~Zhao, ``Quantized space-time
and its influences on some physical problems,'' Chin.\ Phys.\ C \textbf{38},
045102 (2014) doi:10.1088/1674-1137/38/4/045102 [arXiv:1305.3363 [gr-qc]].




\bibitem {INT-GUP2-Mu:2015qta}B.~Mu, P.~Wang and H.~Yang, ``Minimal Length
Effects on Tunnelling from Spherically Symmetric Black Holes,'' Adv.\ High
Energy Phys.\ \textbf{2015}, 898916 (2015) doi:10.1155/2015/898916
[arXiv:1501.06025 [gr-qc]].




\bibitem {INT-GUP2-Benrong:2014woa}B.~Mu, P.~Wang and H.~Yang, ``Covariant GUP
Deformed Hamilton-Jacobi Method,'' Adv.\ High Energy Phys.\ \textbf{2017},
3191839 (2017) doi:10.1155/2017/3191839 [arXiv:1408.5055 [gr-qc]].




\bibitem {INT-GUP2-Chen:2016ftz}L.~Chen and H.~Cheng, ``The tunneling
radiation of a black hole with a $f(R)$ global monopole under generalized
uncertainty principle,'' Gen.\ Rel.\ Grav.\ \textbf{50}, no. 3, 26 (2018)
doi:10.1007/s10714-018-2346-8 [arXiv:1607.07138 [hep-th]].


\bibitem {INT-GUP2-Haldar:2017viz}S.~Haldar, C.~Corda and S.~Chakraborty,
``Tunnelling mechanism in non-commutative space with generalized uncertainty
principle and Bohr-like black hole,'' Adv.\ High Energy Phys.\ \textbf{2018},
9851598 (2018) doi:10.1155/2018/9851598 [arXiv:1705.08307 [gr-qc]].


\bibitem {INT-GUP2-Ong:2018syk}Y.~C.~Ong, ``An effective black hole remnant
via infinite evaporation time due to generalized uncertainty principle,'' JHEP
\textbf{1810}, 195 (2018) doi:10.1007/JHEP10(2018)195 [arXiv:1806.03691
[gr-qc]].


\bibitem {INT-GUP2-Ong:2018xna}Y.~C.~Ong, ``GUP-Corrected Black Hole
Thermodynamics and the Maximum Force Conjecture,'' Phys.\ Lett.\ B
\textbf{785}, 217 (2018) doi:10.1016/j.physletb.2018.08.065 [arXiv:1809.00442
[gr-qc]].


\bibitem {INT-GUP2-Hassanabadi:2019eol}H.~Hassanabadi, E.~Maghsoodi and
W.~S.~Chung, ``Analysis of black hole thermodynamics with a new higher order
generalized uncertainty principle,'' Eur.\ Phys.\ J.\ C \textbf{79}, no. 4,
358 (2019). doi:10.1140/epjc/s10052-019-6871-8




\bibitem {INT-GUP-Faizal:2018fmz}M.~Faizal, S.~E.~Korenblit, A.~V.~Sinitskaya
and S.~Upadhyay, ``Corrections to Scattering Processes due to Minimal
Measurable Length,'' Phys.\ Lett.\ B \textbf{794}, 1 (2019)
doi:10.1016/j.physletb.2019.05.007 [arXiv:1808.01894 [hep-th]].




\bibitem {INT-GUP-Faizal:2017dlb}M.~Faizal, A.~F.~Ali and A.~Nassar,
``Generalized uncertainty principle as a consequence of the effective field
theory,'' Phys.\ Lett.\ B \textbf{765}, 238 (2017)
doi:10.1016/j.physletb.2016.11.054 [arXiv:1701.00341 [hep-th]].




\bibitem {INT-GUP-Faizal:2014mba}M.~Faizal, M.~M.~Khalil and S.~Das, ``Time
Crystals from Minimum Time Uncertainty,'' Eur.\ Phys.\ J.\ C \textbf{76}, no.
1, 30 (2016) doi:10.1140/epjc/s10052-016-3884-4 [arXiv:1501.03111
[physics.gen-ph]].




\bibitem {INT-GUP-Nowakowski:2009ha}M.~Nowakowski and I.~Arraut, ``The Minimum
and Maximum Temperature of Black Body Radiation,'' Mod.\ Phys.\ Lett.\ A
\textbf{24}, 2133 (2009) doi:10.1142/S0217732309030679 [arXiv:0905.3762
[gr-qc]].




\bibitem {INT-GUP-Khodadi:2018wed}M.~Khodadi, K.~Nozari and F.~Hajkarim, ``On
the viability of Planck scale cosmology with quartessence,''
Eur.\ Phys.\ J.\ C \textbf{78}, no. 9, 716 (2018)
doi:10.1140/epjc/s10052-018-6191-4 [arXiv:1808.08436 [gr-qc]].




\bibitem {INT-GUP-Khodadi:2017eim}M.~Khodadi, K.~Nozari and A.~Hajizadeh,
``Some Astrophysical Aspects of a Schwarzschild Geometry Equipped with a
Minimal Measurable Length,'' Phys.\ Lett.\ B \textbf{770}, 556 (2017)
doi:10.1016/j.physletb.2017.05.016 [arXiv:1702.06357 [gr-qc]].




\bibitem {INT-RB-Gim:2018axz}Y.~Gim and B.~Gwak, ``Charged Black Hole in
Gravity's Rainbow: Violation of Weak Cosmic Censorship,'' Phys.\ Lett.\ B
\textbf{794}, 122 (2019) doi:10.1016/j.physletb.2019.05.039 [arXiv:1808.05943
[gr-qc]].




\bibitem {DHE-1Dolan:2011xt}B.~P.~Dolan, ``Pressure and volume in the first
law of black hole thermodynamics,'' Class.\ Quant.\ Grav.\ \textbf{28}, 235017
(2011) doi:10.1088/0264-9381/28/23/235017 [arXiv:1106.6260 [gr-qc]].


\bibitem {DHE-1Gunasekaran:2012dq}S.~Gunasekaran, R.~B.~Mann and D.~Kubiznak,
``Extended phase space thermodynamics for charged and rotating black holes and
Born-Infeld vacuum polarization,'' JHEP \textbf{1211}, 110 (2012)
doi:10.1007/JHEP11(2012)110 [arXiv:1208.6251 [hep-th]].




\bibitem {DHE-1Cvetic:2010jb}M.~Cvetic, G.~W.~Gibbons, D.~Kubiznak and
C.~N.~Pope, ``Black Hole Enthalpy and an Entropy Inequality for the
Thermodynamic Volume,'' Phys.\ Rev.\ D \textbf{84}, 024037 (2011)
doi:10.1103/PhysRevD.84.024037 [arXiv:1012.2888 [hep-th]].




\bibitem {DHE-1Kastor:2009wy}D.~Kastor, S.~Ray and J.~Traschen, ``Enthalpy and
the Mechanics of AdS Black Holes,'' Class.\ Quant.\ Grav.\ \textbf{26}, 195011
(2009) doi:10.1088/0264-9381/26/19/195011 [arXiv:0904.2765 [hep-th]].




\bibitem {DHE-2Christodoulou:1970wf}D.~Christodoulou, ``Reversible and
irreversible transforations in black hole physics,''
Phys.\ Rev.\ Lett.\ \textbf{25}, 1596 (1970). doi:10.1103/PhysRevLett.25.1596




\bibitem {DHE-2Christodoulou:1972kt}D.~Christodoulou and R.~Ruffini,
``Reversible transformations of a charged black hole,'' Phys.\ Rev.\ D
\textbf{4}, 3552 (1971). doi:10.1103/PhysRevD.4.3552




\bibitem {DHE-2Pellicer:1969cf}R.~Pellicer and R.~J.~Torrence, ``Nonlinear
electrodynamics and general relativity,'' J.\ Math.\ Phys.\ \textbf{10}, 1718
(1969). doi:10.1063/1.1665019




\bibitem {DHJ-1-Brau:1999uv}F.~Brau, ``Minimal length uncertainty relation and
hydrogen atom,'' J.\ Phys.\ A \textbf{32}, 7691 (1999)
doi:10.1088/0305-4470/32/44/308 [quant-ph/9905033].




\bibitem {DHJ-1-Guo:2015ldd}X.~Guo, P.~Wang and H.~Yang, ``The classical limit
of minimal length uncertainty relation: revisit with the Hamilton-Jacobi
method,'' JCAP \textbf{1605}, no. 05, 062 (2016)
doi:10.1088/1475-7516/2016/05/062 [arXiv:1512.03560 [gr-qc]].




\bibitem {DHJ-1-Kempf:1996nk}A.~Kempf and G.~Mangano, ``Minimal length
uncertainty relation and ultraviolet regularization,'' Phys.\ Rev.\ D
\textbf{55}, 7909 (1997) doi:10.1103/PhysRevD.55.7909 [hep-th/9612084].


\bibitem {DHJ-1-Berger:2010pj}M.~S.~Berger and M.~Maziashvili, ``Free particle
wavefunction in light of the minimum-length deformed quantum mechanics and
some of its phenomenological implications,'' Phys.\ Rev.\ D \textbf{84},
044043 (2011) doi:10.1103/PhysRevD.84.044043 [arXiv:1010.2873 [gr-qc]].

\end{thebibliography}
\end{document}